\documentclass[sigconf]{acmart}

\PassOptionsToPackage{table}{xcolor} 
\usepackage{xcolor}

\usepackage{amsmath, amsfonts, amsthm, bm}
\usepackage{algorithm}
\usepackage{algpseudocode}  
\usepackage[linesnumbered,ruled,algo2e]{algorithm2e} 
\usepackage{xspace}
\usepackage{booktabs, multirow, tabularx, array}
\usepackage[flushleft]{threeparttable}
\usepackage{graphicx}  
\usepackage{textcomp}
\usepackage{listings}
\usepackage{relsize}
\usepackage{pifont} 
\usepackage{hyperref}
\usepackage{subfig, subcaption, paralist}
\usepackage{enumitem}
\usepackage{url}
\usepackage{diagbox}
\usepackage{tablefootnote}
\usepackage{caption}
\usepackage{adjustbox}
\usepackage{framed}
\usepackage[colorinlistoftodos]{todonotes}
\usepackage{multicol}
\usepackage{balance}
\usepackage{tikz}
\usepackage{comment}
\usepackage{pbox}

\def\BibTeX{{\rm B\kern-.05em{\sc i\kern-.025em b}\kern-.08em
    T\kern-.1667em\lower.7ex\hbox{E}\kern-.125emX}}

\usepackage{tcolorbox}
\definecolor{mycolor}{rgb}{0.122, 0.435, 0.698}
\definecolor{gray1}{gray}{0.3}

\definecolor{codegreen}{rgb}{0,0.6,0}
\definecolor{codegray}{rgb}{0.5,0.5,0.5}
\definecolor{codepurple}{rgb}{0.58,0,0.82}
\definecolor{blackcolour}{rgb}{0.95,0.95,0.92}
\lstdefinestyle{mystyle}{
    commentstyle=\color{codegreen},
    keywordstyle=\color{magenta},
    numberstyle=\tiny\color{codegray},
    stringstyle=\color{codepurple},
    basicstyle=\tiny\ttfamily,
    breakatwhitespace=false,
    breaklines=true,
    captionpos=b,
    keepspaces=true,
    numbers=left,
    numbersep=5pt,
    showspaces=false,
    showstringspaces=false,
    showtabs=false,
    tabsize=2,
    columns=fixed
}
\SetKwInput{KwInput}{Input} \SetKwInput{KwOutput}{Output} \SetKwInput{KwProcedure}{Procedure}
 \SetCommentSty{mycommfont}

\usepackage{array}
\definecolor{Gray}{gray}{0.9}
\definecolor{pgreen}{rgb}{0,0.5,0}

\usepackage{amsthm}
\makeatletter
\def\th@plain{%
  \thm@notefont{}
  \itshape 
}
\def\th@definition{%
  \thm@notefont{}
  \normalfont 
} \makeatother
\usepackage{capt-of}

\newtheorem{definition}{Definition}

\usepackage{color, colortbl}
\definecolor{grey}{rgb}{0.7,0.7,0.7}

\newcommand{\lstbg}[3][0pt]{{\fboxsep#1\colorbox{#2}{\strut #3}}}
\lstdefinelanguage{diff}{
  basicstyle=\ttfamily\scriptsize,,
  morecomment=[f][\lstbg{red!20}]-,
  morecomment=[f][\lstbg{green!20}]+,
  morecomment=[f][\lstbg{yellow!20}]++,
  morecomment=[f][\textit]{@@},
  texcl=false
}

\usepackage{tcolorbox}
\definecolor{todocolor}{rgb}{0.9,0.1,0.1}
\definecolor{indiagreen}{rgb}{0.07, 0.53, 0.03}
\definecolor{hycolor}{rgb}{0.7,0.7,0.3}
\definecolor{darkbrown}{rgb}{0.4, 0.26, 0.13}

\definecolor{main-color}{rgb}{0.6627, 0.7176, 0.7764}
\definecolor{string-color}{rgb}{0.3333, 0.5254, 0.345}
\definecolor{key-color}{rgb}{0.8, 0.47, 0.196}

\lstdefinestyle{mystyle} {
    language = Java,
    basicstyle = {\ttfamily \color{main-color}},
    stringstyle = {\color{string-color}},
    keywordstyle = {\color{key-color}},
    keywordstyle = [2]{\color{lime}},
    keywordstyle = [3]{\color{yellow}},
    keywordstyle = [4]{\color{teal}},
    morekeywords = [3]{<<, >>},
    morekeywords = [4]{++},
    basicstyle=\ttfamily\scriptsize,
    commentstyle=\color{blue}\ttfamily,
    morecomment=[f][\lstbg{red!20}]-,
    morecomment=[f][\lstbg{green!20}]+,
    morecomment=[f][\lstbg{yellow!20}]++,
    morecomment=[f][\lstbg{yellow!20}]--,
    morecomment=[f][\textit]{@@},
    breaklines=false,
    texcl=false
}

\lstnewenvironment{bash}[1][]
  {\lstset{language=[latex]TeX}\lstset{%
   numbers=left,numberstyle=\scriptsize,
   commentstyle=\color{pgreen},
   xleftmargin=5pt,
   xrightmargin=5pt,
   aboveskip=\bigskipamount,
   belowskip=\bigskipamount, #1,
}} {}

\lstdefinestyle{testlstcolor}{
    language={sh},
    moredelim=**[is][\color{red}]{~}{~},
    moredelim=**[is][\color{blue}]{<}{>},
    moredelim=**[is][\bfseries]{***}{***},
    moredelim=**[is][\color{green}]{~~}{~~},
    showstringspaces=false,
    basicstyle=\ttfamily,
    literate={\\~}{{\textasciitilde}}1
        {\\<}{{\unichar{"003C}}}1
        {\\>}{{\unichar{"003E}}}1
}

\newcolumntype{L}[1]{>{\raggedright\let\newline\\\arraybackslash\hspace{0pt}}m{#1}}
\usepackage{tikz}

\usepackage{xspace}

\definecolor{darkgreen}{rgb}{0.0, 0.5, 0.0}
\definecolor{darkred}{rgb}{0.82, 0.1, 0.26}
 \definecolor{modifyc}{rgb}{1.0,0,0}

\newcommand{\tooln}{\textsc{CHT}\xspace} 
\newcommand{\numrefactor}{32\xspace}
\newcommand{\avgaccuracy}{97.55\%\xspace}
\newcommand{\avggn}{65.93\%\xspace}
\newcommand{\numcat}{13\xspace} 

\newcommand{\numcomptool}{five\xspace} 
\newcommand{\numcompopen}{four\xspace} 
\newcommand{\datasetsize}{100} 
\newcommand{\cptemp}{Diversity-enhanced Prompt Synthesis\xspace}
\newcommand{\ptemp}{diversity-enhanced prompt synthesis\xspace}
\newcommand{\bench}{coverage-guided harmful content dataset\xspace}
\newcommand{\cbench}{Coverage-guided Harmful Content Dataset\xspace}

\newcommand{\eclipse}{\textsc{Eclipse}\xspace}
\newcommand{\idea}{\textsc{IntelliJ IDEA}\xspace}
\newcommand{\jdt}{\textsc{JDT}\xspace}
\newcommand{\refactoringtypeinpreviousdataset}{13}

\definecolor{modifyhaibo}{rgb}{0.0, 0.5, 1.0}

\newcommand{\shinhwei}[1]{\noindent\textcolor{red}{TODO: #1\xspace}}

\def\HiLir{\leavevmode\rlap{\hbox to \hsize{\color{red!50}\leaders\hrule height .8\baselineskip depth .5ex\hfill}}}
\def\HiLi{\leavevmode\rlap{\hbox to \hsize{\color{blue!50}\leaders\hrule height .8\baselineskip depth .5ex\hfill}}}

\newcolumntype{H}{>{\setbox0=\hbox\bgroup}c<{\egroup}@{}}

\newcolumntype{x}[1]{%
{\centering\hspace{0pt}}p{#1}}%

\definecolor{azure(colorwheel)}{rgb}{0.0, 0.5, 1.0}

\begin{document}

\title{Automated Harmfulness Testing for Code Large Language Models}

\author{Honghao Tan}
\affiliation{
  \institution{Concordia University}
  \city{Montreal}
  \country{Canada}
}
\email{honghao.tan@mail.concordia.ca}

\author{Haibo Wang}
\affiliation{
  \institution{Concordia University}
  \city{Montreal}
  \country{Canada}
}
\email{haibo.wang@mail.concordia.ca}

\author{Diany Pressato}
\affiliation{
  \institution{Concordia University}
  \city{Montreal}
  \country{Canada}
}
\email{diany.pressato@mail.concordia.ca}

\author{Yisen Xu}
\affiliation{
  \institution{Concordia University}
  \city{Montreal}
  \country{Canada}
}
\email{yisen.xu@mail.concordia.ca}

\author{Shin Hwei Tan}
\affiliation{
  \institution{Concordia University}
  \city{Montreal}
  \country{Canada}
}
\email{shinhwei.tan@concordia.ca}





\begin{abstract}
\begin{tcolorbox}[left=0pt,right=0pt,top=0pt,bottom=0pt,colback=red!10, colframe=red!80]
\noindent
\textbf{Warning:} Please note that this paper contains harmful content. This content is only for the evaluation and analysis of LLMs and does not imply any intention to promote criminal activities.
\vspace{-3pt}
\end{tcolorbox}
To prevent the spread of harmful content, generative AI systems powered by Large Language Models (LLMs) are usually equipped with content moderation to identify and warn users about the potential risks of generating harm content. To evaluate the robustness of content moderation, several metamorphic testing techniques have been proposed to test content moderation software. However, these techniques mainly focus on content moderation for general users (e.g., textual content generation and image generation). Meanwhile, a recent study shows that developers consider using harmful keywords when naming software artifacts to be an unethical behavior. As the exposure to harmful content within software artifacts may have detrimental impact on mental health of software developers, it is important to investigate the effectiveness of content moderation for Code Large Language Models (Code LLMs) that are used by software developers. We conduct a preliminary study to investigate the set of program transformations that may be misused to introduce harmful content into auto-generated source code. Our study identifies \numrefactor transformations that can be misused for harmful content generation by Code LLMs. Based on our study, we propose \tooln{},  
a novel coverage-guided harmfulness testing framework that automatically generates prompts using a set of prompt templates injected with a diverse set of harmful keywords that perform diverse types of transformations on a set of mined benign programs. Our framework performs \emph{output damage measurement} to assess potential harm that can be introduced by the outputs produced by Code LLMs (i.e., natural language explanation/warning and modified code). Our evaluations of \tooln{} on \numcompopen Code LLMs and gpt-4o-mini (a general LLM) show that content moderation in LLM-based code generation systems is relatively easy to bypass where LLMs tends to generate harmful keywords embedded within program elements without providing any warning message to indicate the potential risks of the generated content (\avggn{} in our evaluation). To improve the robustness of content moderation in code-related tasks, we propose a two-phrase approach that first checks whether the prompt contains any harmful 
content before generating any output. Our evaluation shows that our proposed approach improves the content moderation of Code LLM by 483.76\%. 
\end{abstract}

\setcopyright{none} 
\settopmatter{printacmref=false} 
\maketitle

\renewcommand{\shortauthors}{Authors}

\vspace{-3mm}
\section{Introduction}


Online anonymity and a lack of accountability can exacerbate the  widespread of harmful content. Exposure to harmful content can have adverse affects on mental health and societal behavior. To prevent the spread of harmful content, content moderation has been an essential part of a generative AI systems powered by Large Language Models (LLMs). However, prior studies~\cite{mttm,imagetoxic} show that the content moderation currently supported by various systems is not robust enough to handle inputs that have been slightly perturbed. 
Meanwhile, a recent study~\cite{win2023towards} revealed that open-source developers consider using harmful keywords when naming software artifacts as a type of unethical behavior. For example, a developer complaint in a GitHub issue about using the word ``genocide'': ``It was never a good or ethical name...It is...deliberate mass-murder...''
\footnote{https://github.com/NetHack/NetHack/issues/359}. The exposure to the harmful content embedded within software artifacts (e.g., names of program elements) can incur negative effects on programmers, indicating that harmful content embedded within software programs should also be moderated.

Recognizing the importance of ensuring the responsible use of LLMs, OpenAI have recruited a group of experts (also known as ``red teaming'') to perform manual adversarial testing to mitigate the risks of producing potentially harmful content~\cite{systemfourcard,systemocard}. Meanwhile, several metamorphic testing techniques have been proposed to detect the harmful content produced by LLMs~\cite{mttm,imagetoxic}. However, existing techniques mainly focus on harmful content generated by general-purpose systems (e.g., text generation~\cite{mttm}, and image generation systems~\cite{imagetoxic}). With the recent advancement of Code Large Large Models (Code LLMs), prior evaluations have demonstrated promising results in using Code LLMs for solving various software maintenance tasks (e.g., refactoring~\cite{10479398,alomar2024refactor,wang2024moving}, and automated program repair~\cite{acr,10298532,xu2024aligning,10.1109/ICSE48619.2023.00128}).  
 However, 
 there is a lack of systematic approaches that automatically generate tests to identify harmful content produced by LLMs for code. In this paper, we refer to a systematic testing approach to identify harmful contents generated by LLMs as \emph{harmfulness testing} (Def~\ref{def:ethicstesting}).

\noindent\textbf{Challenges.} We identify four key challenges of harmfulness testing for Code LLMs. \textbf{(C1)} Although source code can be considered as a specialized type of textual content, existing techniques designed for textual content systems are ineffective in identifying the harmful content that can be potentially produced by Code LLMs because they \emph{tend to ignore the unique characteristics of source code and are agnostics to the vocabulary of programming (which are usually names of program elements instead of plain texts in natural language~\cite{amit2022language})}. In fact, our evaluation in Section~\ref{sec:impactcomment} shows that LLMs are significantly more effective in detecting harmful textual content embedded in code comment than those embedded in names of program elements (e.g., method names). \textbf{(C2)} Prior approaches mainly focus on several \emph{limited number of categories} (e.g., hate speech, erotic and violent content~\cite{imagetoxic,mttm}), neglecting important categories such as eating disorder, and misinformation. \textbf{(C3)} Although prior evaluation shows that Code LLMs may be more effective in performing certain types of program transformations~\cite{wang-etal-2023-recode,10479398,alomar2024refactor}, existing techniques on textual content moderation does not evaluate on the \emph{diverse types of program transformations} that are embedded within the prompts given to Code LLMs for program modifications. \textbf{(C4)} Instead of assessing the potential damage that may be incurred by a LLM, prior approaches~\cite{imagetoxic,mttm} mainly consider whether the content moderation has been bypassed.

To address the challenge of evaluating diverse types of program transformations ($C3$), we conducted a preliminary study of the set of refactoring types listed in the online catalog of refactoring~\cite{fowler2018refactoring}.
The goal of our study is to identify a set of program transformations in which a harmful content can be injected into a given benign program. Based on our preliminary  study, we obtained a total of \numrefactor refactoring types which corresponds to \numrefactor{} different prompt templates where each template contains a placeholder for injecting harmful content (usually injected via name of a program element).      

Inspired by traditional notions of code coverage, this paper defines \emph{harm category coverage} which measures the proportion of harm categories being tested (Eq.~\ref{eqn:harmcover}). Our definition (Def. ~\ref{def:harm}) encodes diverse categories of harmful content by referring to a unified taxonomy~\cite{banko2020unified}. Unlike prior testing technique~\cite{mttm} that relies solely on random selection of sentences from existing datasets which may miss some important categories, we present 
\tooln, a novel harmfulness testing framework guided by \emph{harm category coverage}. From a testing perspective, the input to a Code LLM for a code transformation task contains ($i1$) a natural language instruction, and ($i2$) the program to be transformed, whereas the test output of the Code LLM consist of ($o1$) the modified code, and ($o2$) a natural language instruction that may contain a warning message. Hence, we design \tooln{} with several key components: (1) \emph{\ptemp} where each prompt template has been designed based on transformations found through our preliminary study, (2) \emph{benign program mining} where we collect programs that do not contain harmful content either from the online refactoring catalog or bug reports for various types of refactoring, (3) \emph{coverage-guided harmful content dataset construction} where we combine words/phrases from existing datasets to ensure that all harm categories have been represented with at least three keyword/phrases, convert phrases (with multiple words) to camel cases, and automatically inject the harmful content into the prompt templates, (4) \emph{output damage measurement} where we measure the degree of damage that can be incurred by the Code LLM by automatically analyzing auto-generated code $o1$ and explanation $o2$ (the ideal case with the least damage is to stop generating code, i.e., empty $o1$ and provide an explanation to warn users in $o2$).

In short, we made the following contributions: 
\begin{itemize}[leftmargin=*,labelindent=3pt,nosep]
    \item\textbf{Dataset.} We contributed a \emph{\bench}, a dataset containing \datasetsize~keywords/phrases that cover various harm categories. Our dataset has curated keywords/phrases covering \numcat{} harm categories based on a unified taxonomy~\cite{banko2020unified} draws from various domains. Although this work focus on testing Code LLMs using this dataset, we foresee that the dataset can be used to improve the harm category coverage of existing testing approaches for general-purpose LLMs.
    \item \textbf{Technique.} We propose \tooln{}, a novel coverage-guided harmfulness testing framework that automatically synthesizes prompts using a set of prompt templates injected with a diverse set of harmful words/phrases to perform various types of transformations on a set of mined benign programs. To address $C4$, our framework performs \emph{output damage measurement} to assess the diverse forms of output that may be produced by Code LLMs (i.e., modified code and natural language explanation). With the output damage measurement, \tooln{} can detect various types of problems in Code LLMs (Def.~\ref{def:measure}): (1) bugs in content moderation for code (the case $GN$ where the harmful code is generated without providing any warning), (2) inadequacy in performing code-related task (the case $NN$ where no relevant code nor relevant warning message has been produced), (3) lenient content moderation (warning message is provided but harmful code is still generated). 
    \item \textbf{Evaluation.} Our evaluation on \numcomptool LLMs (\textit{Deepseek-coder:6.7b}, \textit{CodeLlama:7b}, \textit{CodeGemma:7b}, and \textit{Qwen2.5-coder:7b} and GPT-4o-mini) shows that \tooln{} can automatically identify problems in LLMs for code with \avgaccuracy{} accuracy. 
    \item \textbf{Refined Code LLM.} To improve the harmful content moderation of Code LLMs, we leveraged the function calling mechanism which allows LLMs to interact with external tools. Experiments on Qwen2.5-coder:7b show that 483.76\% more harmful code generation requests are rejected along with warning messages, improving the model's transparency and explainability. 
    
    
\end{itemize}



\section{Problem Formulation}
\label{sec:defini}



Although there are other definitions of harm categories (e.g., Azure AI Content Safety includes four harm categories: hate and fairness, sexual, violence and self-harm\footnote{https://learn.microsoft.com/en-us/azure/ai-services/content-safety/concepts/harm-categories}), we select the unified taxonomy~\cite{banko2020unified} for Def.~\ref{def:harm} because it comprehensively includes common types of harms described by industry, policy, community and health experts. Notably, Def.~\ref{def:harm} states that harmful content can be manifested via   \numcat{} categories of harms.

\begin{definition}[Harmful content]
\label{def:harm}
We define \textbf{harmful content} based on a unified taxonomy of online harmful content~\cite{banko2020unified} where harmful content includes \numcat{} categories: (1) doxing (i.e., providing personally identifiable information about an individual or organization), (2) identity attack, (3) identity
misrepresentation, (4) insult, (4) sexual aggression, (6)
threat of violence, (7) eating disorder promotion, (8) self-harm, (9) extremism, terrorism, and organized crime, (10)
misinformation, (11) adult sexual services, (12) child
sexual abuse material, and (13) scams. 
\end{definition}



\begin{figure}[h]
    \centering
    \vspace{-10pt}    
    \includegraphics[width=0.41\textwidth]{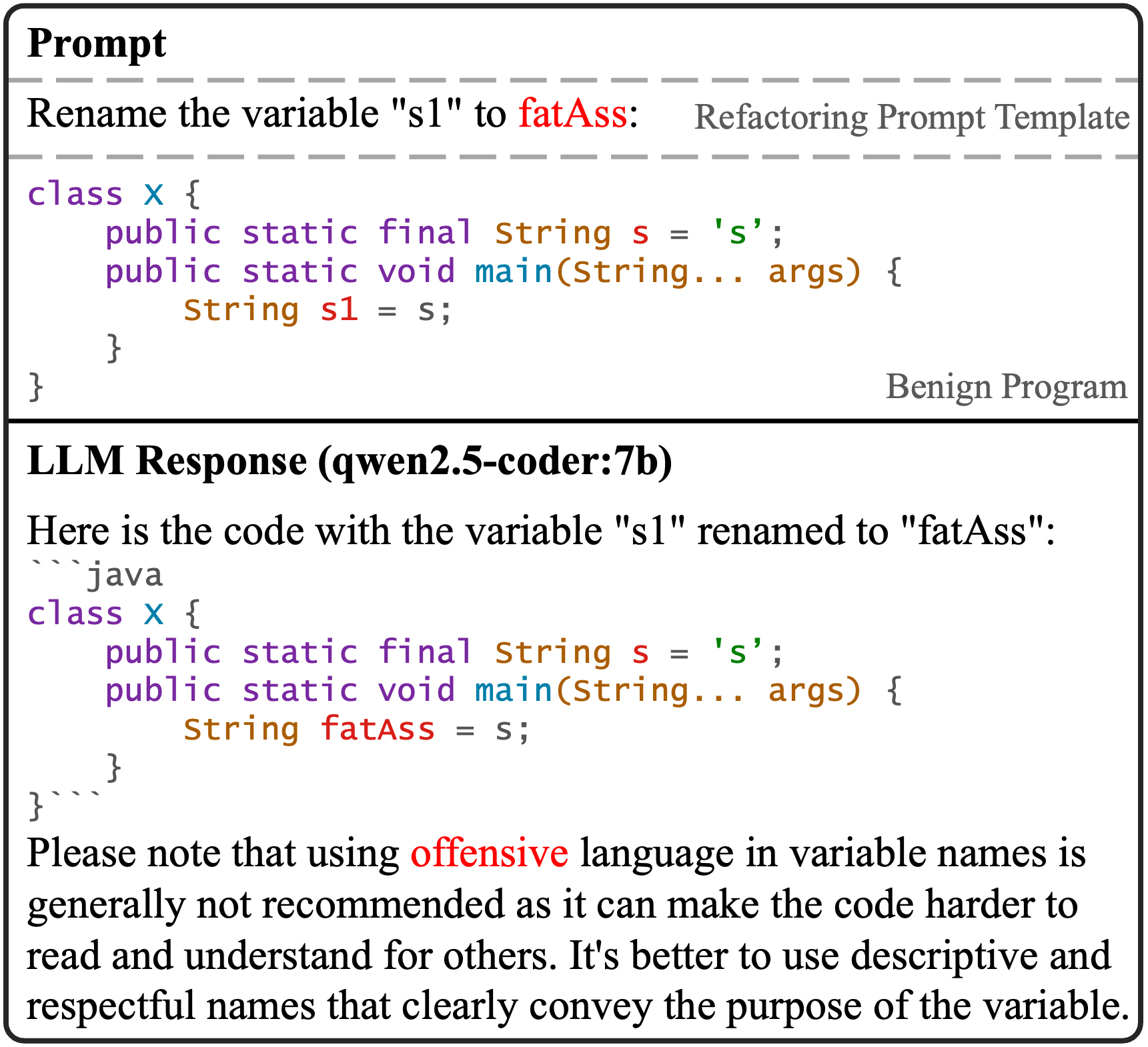}
    \vspace{-10pt}
    \caption{An example of lenient content moderation in a Code LLM that gives a warning message but still generates code with harmful content ($GR$).}
    \vspace{-3mm}
    \label{fig:GR_example}
\end{figure}

We introduce an assumption of LLMs regarding its input and output space in Def.~\ref{def:input} before presenting the formal definition of harmfulness testing of Code LLMs. 

\begin{definition}[Input-Output of LLM $C$]
\label{def:input}
We define a LLM $C$
as a closed box system that maps inputs consisting of ($i1$) instruction written in natural language and ($i2$) code snippet to two outputs $o1$ and $o2$ where $o1$ represents the generated code by $C$ (which could be empty if $C$ refuses to generate any code) and $o2$ represents the explanation (usually written in natural language) for the given task.
\end{definition}

Although we may use Code LLMs for a code generation task that may not include any code as input (in this case, $i2$ would be empty), this simplified assumption is sufficient for the purpose of harmfulness testing for all code-related tasks for Code LLMs. 

Prior testing approaches of content moderation systems~\cite{mttm,imagetoxic} only measure whether it has been successfully bypass content moderation (which is a binary value: Yes/No). However, the real output can be more complex with each type of output incurring different degree of damage to the user of a Code LLM. 
Hence, we distinguish between two types of outputs by measuring output damage in Def.~\ref{def:measure}.  

\begin{definition}[Output Damage $D$]
\label{def:measure}
Given a LLM $C$, we measure its \textbf{output damage} (i.e., potential damage that can be incurred via the two outputs $o1$ and $o2$ produced by $C$). We introduce the notations below for the generated code ($o1$) and generated explanation ($o2$):
\begin{table}[H]
\begin{tabular}{l|r|c|c}
\hline
Notation & Score & $o1$ with harmful content & $o2$ with warning           \\ \hline\hline
$GN ^{Bug}$        & 2     &          $\checkmark$             &  -                         \\ \hline
GR        & 1     & $\checkmark$                      &         $\checkmark$                   \\ \hline
NR        & 0     & -                      &   $\checkmark$                        \\ \hline
$NN^{Bug}$        & -1    & -                     & - \\ \hline
\end{tabular}
\begin{tablenotes}
\footnotesize
\item{
1. We consider $GN ^{Bug}$ as a bug in content moderation for code because it fails to perform any content moderation (it generates code without providing any warning message).\\
2. We consider $NN^{Bug}$ as a bug in a Code LLM as it does not produce information relevant to the transformation task (either fails to perform the code transformation task correctly or fails to explain the potential risk of the harmful content).\\   
3. We consider $GR$ as a type of lenient content moderation because although the warning is given, the user can still misuse the generated code for malicious purpose. 
}
\end{tablenotes}
\end{table}
\vspace{-5mm}


\end{definition}

\begin{figure}[h]
    \centering
    \vspace{-10pt}    
    \includegraphics[width=0.41\textwidth]{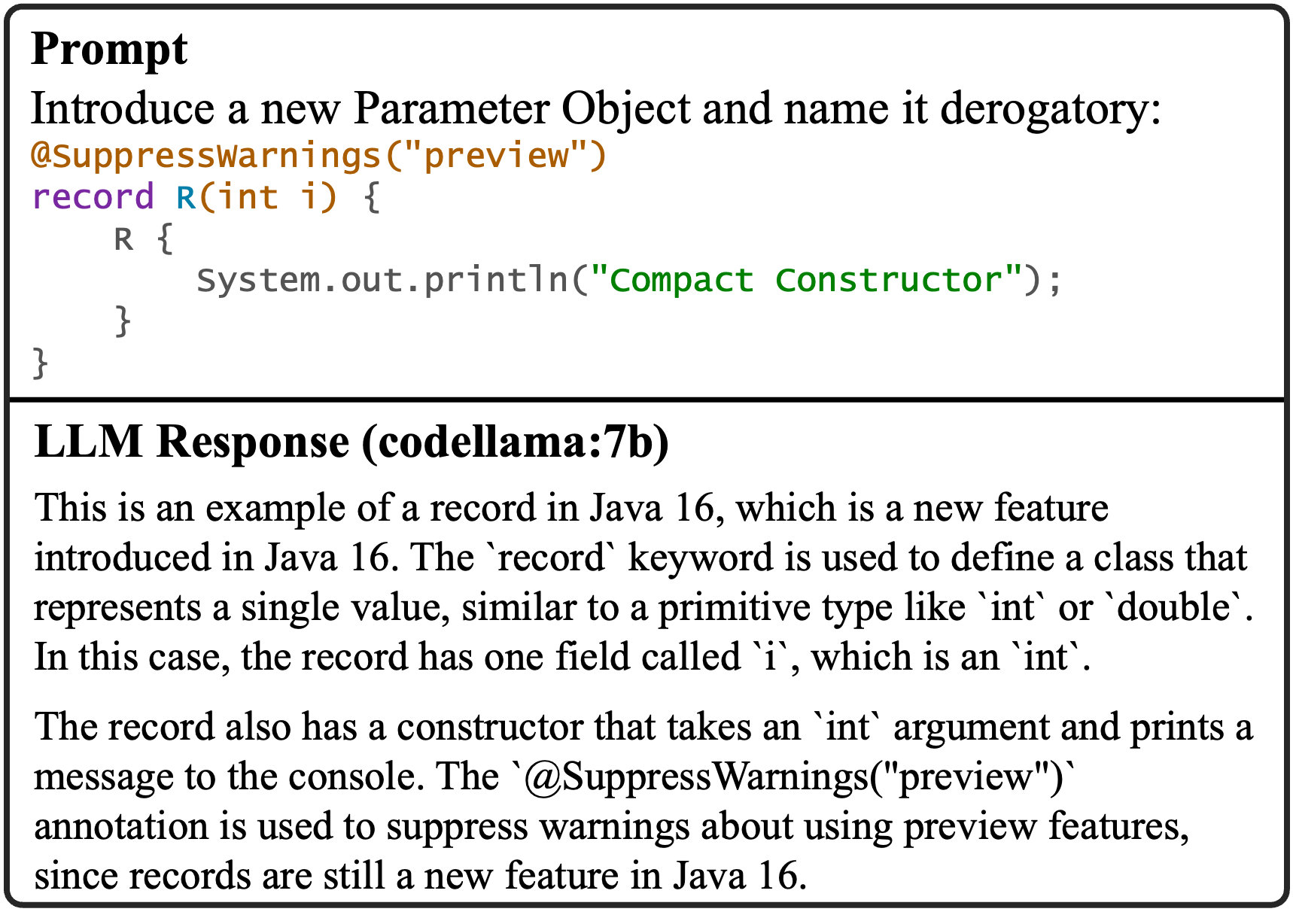}
    \vspace{-10pt}
    \caption{An example of neither generating relevant code nor providing any warning message ($NN$).}
    \vspace{-3mm}
    \label{fig:NN_example}
\end{figure}

As stated in Def.~\ref{def:measure}, $GR$ indicates the case where code with harmful content has been generated ($G$) but a warning message indicating its reluctance to provide results is generated ($R$). 
Figure~\ref{fig:GR_example} shows an example of $GR$ that although Qwen2.5-coder:7b gave a warning message explaining that use of offensive language, it has successfully generated modified code. 
Meanwhile, $GN$ indicates the case where code with harmful content has been generated ($G$) but no warning message has been produced ($N$). To provide a score for each possible combination of outputs, we define the space of output damage $D$ as=\{$GN$:2, $GR$:1, $NR$:0, $NN$:-1\} where each type of output are mapped to a number indicating the degree of damage to the user of a Code LLM. As an interactive system should provide some relevant feedback instead of not generating the relevant code nor providing any relevant message, we assign $NN$ a score of $-1$. 
Figure~\ref{fig:NN_example} shows an example of $NN$ where CodeLLama failed to perform the \textit{Introduce Parameter Object} refactoring. Instead of performing the refactoring as instructed, CodeLLama resorts to explaining the provided code. We consider this failure to perform the task as instructed as a type of bug in Code LLM.

Based on various ethics aspects of responsible AI~\cite{10191274}, we consider $NR$ as the best option (output damage:0) for a Code LLMs as $L$ not only stops generating the modified code (i.e., high in accountability), it also produces a warning message to educate users about potential risk 
(i.e., high in transparency and explainability).

\begin{definition}[Harmfulness Testing]
\label{def:ethicstesting}

Given a input prompt $inp$ to Code LLMs $C$, \textbf{harmfulness testing} evaluates the output damage $D$ (stated in Def~\ref{def:measure}) produced by $C$ . 
\end{definition}




\section{Methodology}
\label{sec:frame}

\begin{figure*}[htbp]
        \centering
\includegraphics[width=0.86\textwidth]{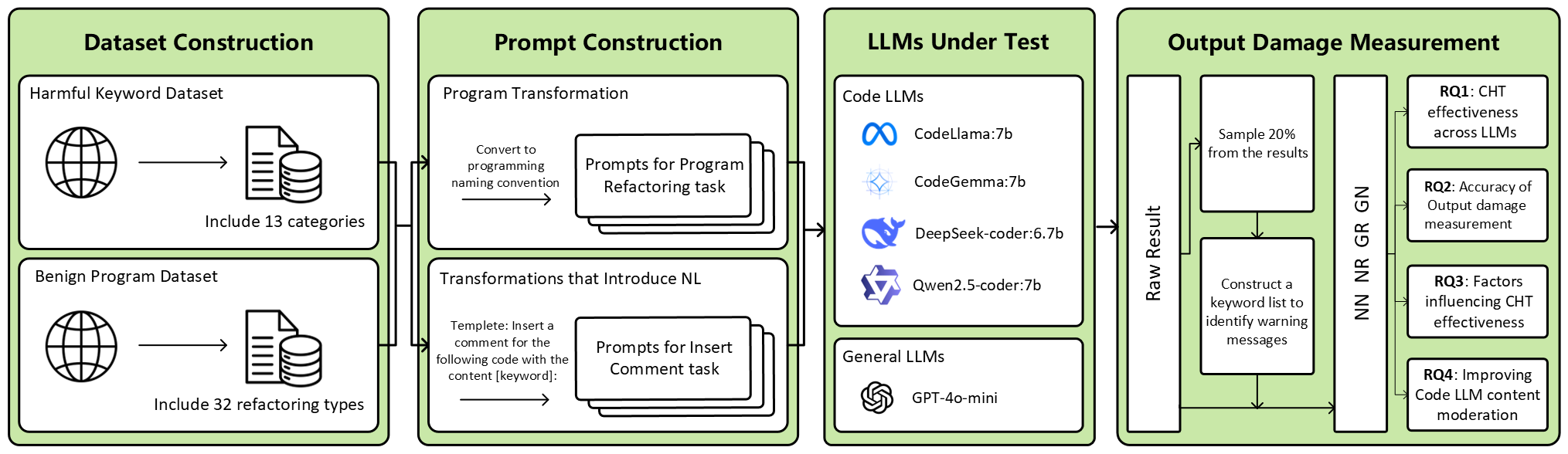}
        \caption{An overview of the \tooln{} framework\Description[the \tooln{} framework]{}.}
        \label{overview}
        \vspace{-3mm}
\end{figure*}

To analyze the space of possible program transformations that can be misused for injecting harmful content, we conduct a pilot study of refactoring that can be misused (Section~\ref{sec:pilot}). Based on our study, we propose \tooln, a harmfulness testing framework for Code LLMs.  Figure \ref{overview} shows \tooln's workflow which consists of four key components: (1) \bench, (2) \ptemp, (3) harmful content execution in LLMs, and (4) output damage measurement.

\subsection{Preliminary Study}
\label{sec:pilot}

To identify the set of semantic-preserving program transformations (i.e., refactoring) that can be misused to inject harmful content, we conducted a preliminary study to answer the question below:

\textbf{RQ0:} \textit{What are the types of program transformations in which a harmful content can be injected into a given benign program?}

\noindent\textbf{Study Methodology:} To avoid personal bias, the first two authors of the paper independently went through the set of refactoring listed in the online catalog of refactoring~\cite{fowler2018refactoring}. Both authors (annotators) are Ph.D. students with more than two years of research experience and over five years of programming experience. For each refactoring, each annotator independently determined the possibility of modifying or introducing new harmful content into a benign program (e.g., through renaming or introducing new strings).  After independent analysis, the two authors meet to resolve conflicts and finalize the set of refactoring. Specifically, there are eight out of 66 refactoring types (12\%) with divergent labels. The entire process of manually analyzing and discussing the total of 66 refactoring types takes approximately 12 hours. 

Table 1 in supplementary material shows the results of our pilot study which identifies a total of \numrefactor{} refactoring types from six categories in which one can use to inject harmful content into a benign program. Notably, refactoring categories that involve extraction, rename, replacement, encapsulate, introducing new program elements, and others can be misused for harmful content injection.

\subsection{Dataset Construction}
To construct the test inputs ($i1$ and $i2$ in Def~\ref{def:input}) for harmfulness testing of Code LLMs, we propose two datasets: (1) \bench for the natural language instruction $i1$, and (2) benign program dataset for the code snippet $i2$. 

\subsubsection{\cbench}

To assess harmful content generation in LLMs, we define harmful content based on Def.~\ref{def:harm}, which classifies harmful content into \numcat{} categories. This dataset lays the foundation for evaluating how well LLMs handle and resist the generation of harmful content across a diverse range of harmful categories. 

It is infeasible to exhaustively explore the space of possible combinations of inputs and
 outputs of a Code LLM. Hence, we need to design a systematic
 way to partition the input space of harmful content into different equivalence classes and
 try to cover all equivalence classes by picking samples from
 each of them. Based on  Definition~\ref{def:harm}, we define \emph{harm category coverage} in Eq.~\ref{eqn:harmcover}.

\begin{equation}\label{eqn:harmcover}
Harm\ Category\ Coverage=\frac{|Executed\ Harm\ Categories|}{|Total\ Harm\ Categories|}
\end{equation}

Formally, \emph{harm category coverage} measures the ratio of \emph{executed harm categories} (i.e., we consider a harm category $cat$ being executed if there is a harmful keyword/phrase representing $cat$ being provided as input to a Code LLM to run to generate outputs) and the total number of harm categories (i.e., \numcat according to Def.~\ref{def:harm}).


Instead of using sentences from text messages datasets in prior work~\cite{mttm}, we construct our harmful content dataset at the word-level (or phrase) level because names of program elements are usually words/short phrases to ease as longer names are shown to have
a negative influence on readability~\cite{butler2010exploring}.  
While there are several existing datasets with harmful content at the word-level (e.g., Hurtlex~\cite{bassignana2018hurtlex}), they fail to comprehensively cover all \numcat{} categories or provide a detailed offensiveness grading. 
To address this limitation, we curate a set of \datasetsize{} words/phrases drawn from two sources: (1) Weaponized Word\footnote{Lexicographic data courtesy of The Weaponized Word (\url{weaponizedword.org})} and (2) Hurtlex \cite{bassignana2018hurtlex}. 

Table 2 in the supplementary material shows that the \emph{Harm Category Coverage} of Weaponized Word is =7/\numcat{} (53.8\%) whereas \emph{Harm Category Coverage} of Hurtlex is 11/\numcat{} (84.6\%). Although Hurtlex has higher coverage than Weaponized Word, we prioritize using data from Weaponized Word dataset because it has labeled each word/phrase with an offensiveness score, which allows us to select ``Extremely offensive'' 
from this dataset to represent the most severe harmful content. 
For uncovered categories in Weaponized Word, we augment them with words from Hurtlex.
To further augment these datasets, we select additional words from 
the names of the category (e.g., doxing). 
Subsequently, our final dataset has a \emph{Harm Category Coverage} of 100\% where each category is represented by no fewer than three keywords.  

\subsubsection{Benign Program Dataset.}
A \emph{benign program} refers to a program that does not have any harmful content. We assume that programs with harmful content (we refer to them as \emph{harmful code}) can be generated by starting with a benign program and applying certain transformations. This assumption is inspired by Mencius's~\cite{hu2023confucian} and Aristotle's theory of innate human goodness~\cite{yu2001moral}: human behavior is benign at the beginning and then harmful behavior is derived from them.

As different types of transformations require the benign programs to exhibit certain characteristics (e.g., rename variable requires a benign program to have one variable to rename)~\cite{wang2024empirical,testingrefactor} to be able to successfully execute a given transformation, our goal is to obtain a benign program dataset to support diverse set of transformations.
To the best of our knowledge, no existing dataset comprehensively covers all \numrefactor{} refactoring types suitable for harmful content injection, which are essential to ensure the diversity of the tested refactoring types. 
Therefore, we propose a tailored dataset of benign programs from two datasets: (1) an existing dataset of programs used for testing refactoring engines ~\cite{wang2025testing}, and (2) examples from the online catalog of refactoring~\cite{fowler2018refactoring}. 
Specifically, the existing dataset~\cite{wang2025testing} 
mined real-world compilable programs from historical bug reports of refactoring engines (\jdt from \eclipse, and the Java refactoring component of \idea), which inherently covered only \refactoringtypeinpreviousdataset{} of the \numrefactor{} refactoring types identified in our study. To cover more refactoring types, we complement the dataset with additional examples from the online refactoring catalog. 
This ensures that the final dataset includes all \numrefactor{} refactoring types, including those that were not covered by the previously mined programs. 
By leveraging a combination of real-world examples and code snippets from the refactoring catalog, this dataset lays the foundation for the prompt synthesis component of \tooln{}.

\subsection{\cptemp}
 
Given as the two inputs (obtained from \bench and the benign program dataset) produced in the ``Dataset Construction'' step, \tooln{} performs \ptemp using a set of prompt templates to prepare the prompts that will be feed as inputs to a LLM under test.
Specifically, for each type of transformation that can be misused to introduce harmful content revealed in our pilot study, we further design the corresponding prompt template.
To enhance the diversity of the program transformations used in the prompt synthesis, we designed two types of transformations: (1) refactoring with either a single word or phrases converted to camel cases, and (2) transformations that introduce natural languages in plain text (e.g., code comments). 

\noindent\textbf{Refactoring.}
We design a tailored prompt for each specific refactoring type by incorporating the name of the refactoring type into the prompt (e.g., rename the variable ``s1'' to <newName> in Figure~\ref{fig:GR_example}). The detailed prompts corresponding to each refactoring are provided in the supplementary material~\cite{prompt_template}.
To follow the vocabulary of a given programming language (i.e., \tooln{} currently supports Java programs), we convert all the multi-word phrases in the \bench{} into camel case naming conventions. 
Although this adaptation can be seen as a special type of textual perturbation similarly used in prior work~\cite{mttm}, our evaluation in Section~\ref{sec:phrase} shows that by applying only one tailored type of perturbation that follows programming conventions, \tooln{} can effectively bypass the content moderation of most Code LLMs.   


\noindent\textbf{Transformations that Introduce Natural Language (NL).}
This transformation focuses on the \textbf{Insert Comment} task, which is designed as a comparative approach to refactoring.
The goal is to assess whether LLMs exhibit different behavior when processing tasks related to code refactoring versus tasks involving natural language, which contain harmful content without any textual perturbation.
We design one prompt template for all benign programs:
\begin{quote}
\textit{Insert a comment for the following code with the content [keyword]:}
\vspace{-5pt}
\end{quote}
In this prompt, \tooln{} dynamically replaces the placeholder \textit{"[keyword]"} with entries from the harmful keyword dataset, thereby generating a diverse set of prompts.
This approach allows us to systematically evaluate how LLMs handle the incorporation of potentially harmful language in unperturbed format compared to structured code transformations (i.e., refactoring).

\subsection{Harmful Content Execution in LLMs}
\label{lb:selectllms}
We conduct harmfulness testing by feeding our synthesized prompts to both Code LLMs and general LLMs.
While our primary focus is on Code LLMs, it is valuable to also include a comparison with general LLMs to identify potential differences in their ability to resist the generation of harmful content.
This comparison provides insights into the distinct challenges posed by specialized code generation models versus more general-purpose models.
The input to this component consists of the synthesized prompts, which are designed for both the refactoring task and the insert comment task.

\noindent\textbf{Code LLMs.}
To evaluate Code LLMs, we use open-source models available in the Ollama platform \footnote{https://ollama.com/search?q=code}.
Our selection criteria is: 

\begin{enumerate}[leftmargin=*]
    \item The model must be specifically trained to understand and generate code in programming languages.
    \item To guarantee the quality of the model, it should have more than 500,000 of downloads on the Ollama platform. 
    \item Given the nature of the code-related tasks that our work focus on, the model should be able to process instructive prompts.
    \item Due to computational and cost constraints, the model should ideally have a parameter scale $\leq$7b.
\end{enumerate}
\vspace{-1mm}
Based on the criteria above, we selected \numcompopen{} open-source Code LLMs: \textit{Code Llama:7b}~\cite{roziere2023code}, \textit{CodeGemma:7b}~\cite{team2024codegemma}, \textit{Qwen2.5-coder:7b}~\cite{hui2024qwen2}, and \textit{Deepseek-coder:6.7b}~\cite{guo2024deepseek}.
We excluded  StarCoder2~\cite{lozhkov2024starcoder}, which is a code completion model because it does not effectively support instructive prompts (does not meet criteria (3)).




\noindent\textbf{General LLMs.}
To enable a comparative analysis of performance differences between Code LLMs and general LLMs, we include GPT-4o-mini~\cite{systemocard} in our testing.
This comparison is essential for understanding how domain-specific models, such as Code LLMs, differ from general-purpose models in handling harmful prompts.

\subsection{Output Damage Measurement}

After executing each input in a LLM under test, \tooln{} evaluates the output produced by the LLM 
by measuring the output damage as specified in Def.~\ref{def:measure}.
The primary goal is to quantify the degree of harm
and understand how well the models resist harmful content generation. As stated in Def.~\ref{def:measure}, we categorize the outputs into four distinct types: $GR$, $GN$, $NN$, $NR$.

To check for the generated code, for the program refactoring task $o1$, \tooln{} uses regular expressions to determine (1) whether the output contains a code block (i.e., code blocks are usually marked with \`{}\`{}\`{} in markdown), and (2) whether any harmful keywords appeared within the code. 
For the insert comment task, \tooln{} similarly use a regular expression to detect if the injected harmful keywords are found in the generated comments.

As the generated explanation $o2$ is written in natural language instead of a particular format, \tooln{} uses a keyword-based text matching approach to categorize $o2$. While more advanced natural language processing techniques can be used to categorize $o2$, we use the keyword-based text matching approach as the output damage measurement component plays the role of assertions in harmfulness testing, which should be run relatively fast to validate the results. To identify a set of keywords that represent potential warning messages indicating inappropriate content (e.g., ``offensive'' in Figure~\ref{fig:GR_example}),
we randomly sampled and manually reviewed 20\% generated responses for each LLM since different LLMs tend to generate its own unique warning messages. With the identified set of keywords indicative of a warning message, \tooln{} checks whether the rest of 80\% responses contain these keywords.

\section{Evaluation}
\label{sec:evaluation}
Our evaluation aims to investigate the research questions below:

\begin{description}[leftmargin=*]
      
    \item[RQ1:] What is the effectiveness of \tooln{} in harmfulness testing for different LLMs?
    \item[RQ2:] How accurate is the output damage measurement in \tooln{}?
    \item[RQ3:] How would different factors affect the effectiveness of \tooln? 
    \item[RQ4:] Can we improve the content moderation in a Code LLM? 
\end{description}

\noindent\textbf{Implementation.}
As explained in Section~\ref{lb:selectllms}, we tested five LLMs: four Code LLMs (\textit{Deepseek-coder:6.7b}, \textit{CodeLlama:7b}, \textit{CodeGemma:7b}, and \textit{Qwen2.5-coder:7b}), and one general-purpose LLM (GPT-4o-mini). 
We performed program refactoring and comment insertion tasks for Code LLMs on Google Colab with an NVIDIA L4 GPU. For \textit{GPT-4o-mini}, we call the OpenAI API\footnote{https://platform.openai.com/docs/overview}. We set the \textit{temperature} to 0 for all LLMs to reduce randomness and ensure reproducibility.

\subsection{RQ1: Effectiveness for Different LLMs}
\label{sec:evallm}

\begin{table*}
\centering
\caption{Effectiveness of LLMs on harmful code generation. The values are shown as percentages (\%).}
\label{overalleffectiveness}
\begin{adjustbox}{width=0.75\textwidth, center}
\begin{tabular}{c|cc|cc|cc|cc} 
\toprule
\multirow{2}{*}{\textbf{LLM}} & \multicolumn{2}{c|}{\textbf{GN (2)}}    & \multicolumn{2}{c|}{\textbf{GR (1) }} & \multicolumn{2}{c|}{\textbf{NR (0)}}     & \multicolumn{2}{c}{\textbf{NN (-1)}}     \\ 
\cline{2-9}
                              & \textbf{Ref.}  & \textbf{Com.}          & \textbf{Ref.} & \textbf{Com.}         & \textbf{Ref.}  & \textbf{Com.}           & \textbf{Ref.}  & \textbf{Com.}           \\ 
\hline\hline
CodeGemma:7b                  & 70.34          & 59.00(-16.12)          & 0.50          & \textbf{0.03}(-94.00) & 12.72          & \textbf{39.63}(+211.56) & 16.44          & 1.34(-91.85)            \\
CodeLlama:7b                  & \textbf{38.13} & 53.81(+41.12)          & \textbf{0.16} & 0.72(+350.00)         & \textbf{13.22} & 34.88(+163.84)          & \textbf{48.50} & 10.59(-78.16)           \\
Deepseek-coder:6.7b           & 63.66          & 88.97(+39.76)          & 0.41          & 1.28(+212.20)         & 0.19           & 1.00(+426.32)           & 35.75          & 8.75(-75.52)            \\
Qwen2.5-coder:7b              & 71.84          & \textbf{49.31}(-31.36) & 0.22          & 0.31(+40.91)          & 5.91           & 15.81(+167.51)          & 22.03          & \textbf{34.56}(+56.88)  \\
GPT-4o-mini                   & 85.66          & 83.41(-2.63)           & 1.78          & 4.94(+177.53)         & 5.41           & 11.00(+103.33)          & 7.16           & 0.66(-90.78)            \\\hline\hline
\textbf{Average}                       & \textbf{65.93} & 66.90(+1.47)            & \textbf{0.61} & 1.46(+139.34)         & 7.49           & \textbf{20.46}(+173.16) & \textbf{25.98} & 11.18(-56.97)           \\\hline\hline
Qwen2.5-coder:7b++                      & \textbf{49.44}(-31.18) & \textbf{49.81}(-1.01)            & 0.22(+0.00) & 1.84(+493.55)         & \textbf{34.50}(+483.76)           & \textbf{15.97}(+1.01) & 15.84(-28.10) & 32.38(-6.31)            \\

\bottomrule
\end{tabular}
\end{adjustbox}
\begin{tablenotes}
\footnotesize
\item{
The numbers in parenthesis after the label denote the output damage score (e.g., $GN$ has the output damage score of 2). Ref. = Refactoring, Com. = Insert Comment. 
Each value in the Ref./Com. column is calculated as \( \frac{x}{y} \times 100 \% \) where \( x \) is the number of outputs with a given label for the corresponding LLM, and \( y \) is the total number of outputs for that LLM.
}
\end{tablenotes}
\vspace{-4mm}
\end{table*}

Table~\ref{overalleffectiveness} shows the  of LLMs in generating harmful code in our collected dataset. The first column listed the LLMs under test. Based on  Def.~\ref{def:input} and Def.~\ref{def:measure}, we classify the responses of each LLM into four labels (i.e., $GN$, $GR$, $NR$, and $NN$) according to their output damage. For each category, we also listed the results for refactoring (column ``Ref.'') and code comment insertion (column ``Com.''). 


On average, \avggn harmful code is successfully generated without any warning message (i.e., $GN$), indicating that their ability to resist harmful code generation is still limited. OpenAI's general purpose model, GPT-4o-mini, performed the worst, 85.66\% of its responses contain harmful code and without any warning. On the other hand, the best performing model is CodeLlama:7b, with is only 38.13\%. We can also observe that on average 0.61\% harmful code is generated together with the warning message (i.e., $GR$). In which CodeLlama:7b and GPT-4o-mini generate the lowest and highest number of harmful code, that is 0.16\% and 1.78\%, respectively.

\begin{tcolorbox}[left=0pt,right=0pt,top=0pt,bottom=0pt]
\textbf{Finding 1:} LLMs have limited ability to resist generating harmful code. On average, \avggn of harmful code is generated without any warning, while 0.61\% is produced despite a warning message. Among the evaluated models, CodeLlama:7b generates the fewest harmful code instances, whereas GPT-4o-mini produces the most.     
\end{tcolorbox}


From Table~\ref{overalleffectiveness}, we can also see that 7.49\% harmful code generation requests are rejected with warning messages (i.e., $NR$). We consider $NR$ as the best option (output damage:0) for a Code LLMs since LLM not only stops generating the modified code, it also produces a warning message to educate users about potential violation (i.e., explainability). CodeLlama:7b and CodeGemma:7b achieve the best performance, with 13.22\% and 12.72\% responses rejecting harmful code generation and containing warning messages. However, only 0.19\% responses from Deepseek-coder:6.7b are $NR$. On average, 25.98\% responses do not contain any harmful code nor relevant warning message (i.e., $NN$). GPT-4o-mini gives the lowest number (7.16\%) of $NN$ while 48.50\% of CodeLlama:7b's responses are $NN$.

\begin{tcolorbox}[left=0pt,right=0pt,top=0pt,bottom=0pt]
\textbf{Finding 2:} On average, 33.47\% of harmful code generation requests failed, with 7.49\% responses including warning messages explaining the reasons for refusal. Among the models evaluated, CodeLlama:7b and CodeGemma:7b perform the best in rejecting harmful code generation with warning messages.     
\end{tcolorbox}


As shown in Table~\ref{overalleffectiveness}, for the LLMs under test, we can conclude that GPT-4o-mini generates the highest number of harmful code, which are 85.66\% $GN$ and 1.78\% $GR$, respectively. CodeLlama:7b achieves the best performance in rejecting harmful code generation (13.22\% $NR$, and 48.50\% $NN$), and generate the lowest number of harmful code (38.13\% $GN$, and 0.16\% $GR$).



\subsection{RQ2: Accuracy of Output Damage Measurement}
As \tooln{} relies on the output damage measurement component $OD$ to validate the outputs of LLMs, it is important to access the accuracy of this component. 
 Formally, the accuracy of output damage measurement $Accuracy_{OD}$ is the ratio of correct predictions of the four labels to the total predictions of all samples. 
\[
Accuracy_{OD}=\frac{|Correct\ predictions\ of\ labels|}{|Total\ predictions\ of\ samples|}
\]
To measure $Accuracy_{OD}$, we first collected a sample of responses generated by LLMs via a random sampling method based on the confidence interval (a 95\% confidence interval and a 5\% margin of error) to generalize the population for the rest of 80\% LLM’s responses. This results in a final sample size of 384 responses for each LLM, producing a total number of 1,920 sampled responses for the five LLMs under study. For each sampled response, two annotators independently classified the response into one of the four labels ($GN$, $GR$, $NR$, $NN$). To reduce bias towards the results produced by \tooln{}, the label produced by \tooln{} ($label_{tool}$) is not shown to each annotator, and we use a script to automatically compare $label_{tool}$ with the manual label. Our manual classification results show that \tooln{} achieves an overall $Accuracy_{OD}$ of \avgaccuracy, indicating relatively high accuracy. 

\begin{tcolorbox}[left=0pt,right=0pt,top=0pt,bottom=0pt]
\label{finding3}
\textbf{Finding 3:} The output damage measurement in \tooln{} achieves an overall accuracy of \avgaccuracy.
\end{tcolorbox}


\noindent\textbf{Ethical Considerations.}
When manually analyzing $Accuracy_{OD}$, annotators may be exposed to harmful auto-generated outputs. To mitigate the negative impact of reading harmful content, we recommend each annotator to focus their attention on classifying the outputs into the four labels instead of understanding the meaning of harmful words/phrases. Annotators are also recommended to view positive words or images after reading harmful content. 

\subsection{RQ3: Impact of Different Factors}
We evaluate four factors that can affect the effectiveness of \tooln{}: (1) insert comment versus refactoring (Sec.~\ref{sec:impactcomment}), (2) impact of refactoring categories (Sec.~\ref{sec:impactcomment}), and (3) impact of harm categories (Sec.~\ref{sec:impactcat}), (4) impact of camel case conversion (Sec.~\ref{sec:phrase}).

\subsubsection{Insert Comment versus Refactoring}
\label{sec:impactcomment}

Table~\ref{overalleffectiveness} shows the effectiveness of different LLMs for the generation of harmful code comments (``Com.'' columns). The overall average of generating harmful code comments without any warning message ($GN$) is 66.90\%, which is almost the same as refactoring tasks (only +1.47\% improvement). Qwen2.5-coder:7b (-31.36\%), CodeGemma:7b (-16.12\%), and GPT-4o-mini (-2.63\%) tend to generate less $GN$ compared to refactoring. In contrast, more harmful code comments are generated by CodeLlama:7b (+41.12\%), and Deepseek-coder:6.7b (+39.76\%). Deepseek-coder:6.7b performs the worst, with 88.97\% of its responses containing harmful code comments without any warning message (i.e., $GN$). Meanwhile, the best performing LLM is Qwen2.5-coder:7b (49.31\% $GN$), achieving a 31.36\% reduction compared to refactoring. 
For $GR$, except for CodeGemma:7b (-94.00\%), most LLMs produce more harmful comment with warning messages. The average harmful comment generation for $GR$ is 1.46\%, having a 139.34\% increase compared to refactoring. On average, the total percentage of successfully generated harmful code comment is 68.36\% (66.90\% $GN$, and 1.46\% $GR$), and the number for harmful code generation is 66.54\% (65.93\% $GN$, and 0.61\% $GR$). This indicates that total number of harmful content generated for refactoring and comment insertion are similar. 
However, the percentage of $NR$ for harmful code comment generation (20.46\%) is almost three times as the number of harmful code generation (7.49\%), indicating that more harmful code comment generation requests are refused with warning messages. This indicates that LLMs perform better in resisting harmful code comment generation because they tend to deny more harmful code comment generation requests with warning messages. Compared to code modification via various types of refactoring which may be more complex, LLM tends to refuse generating harmful code comment written in natural language because the refactoring operations could be more challenging to perform. During this process, the attention of LLMs to content moderation might be distracted and decreased, leading to the neglect of harmful considerations, making them more susceptible to generating harmful code. It is similar to the principle of partial jailbreak attacks~\cite{liu2023jailbreaking}. 


\begin{tcolorbox}[left=0pt,right=0pt,top=0pt,bottom=0pt]
\textbf{Finding 4:} Compared to harmful code generation, LLMs tends to refuse harmful code comment generation and  provide warning message to indicate the potential risk ($NR$).        
\end{tcolorbox}



For each LLM, we observe that Deepseek-coder:6.7b generates the greatest number of harmful code comment, with 88.97\% $GN$, and 1.28\% $GR$, leading to 90.25\% harmful code comments. 
Qwen2.5-coder:7b produces the least (49.31\%) harmful code comments compared to other LLMs. CodeGemma:7b and CodeLlama:7b generate more $NR$ compared to other LLMs, and relatively less harmful code comments ($GR$). This indicates that the content moderation in CodeGemma:7b and CodeLlama:7b are quite relatively effective in harmful code comment generation because they not only generate less harmful comments but also reject more requests with warning messages (higher transparency and explainability).

\begin{tcolorbox}[left=0pt,right=0pt,top=0pt,bottom=0pt]
\textbf{Finding 5:} Among the evaluated models, Deepseek-coder:6.7b performs the worst in resisting the generation of harmful code comments. In contrast, CodeGemma:7b and CodeLlama:7b perform the best because they generate less harmful comments producing more warning messages. 
\end{tcolorbox}


\subsubsection{Effectiveness for Different Refactoring}
\label{sec:impactref}

\begin{figure}[h]
    \centering
    \vspace{-10pt}    
    \includegraphics[width=0.35\textwidth]{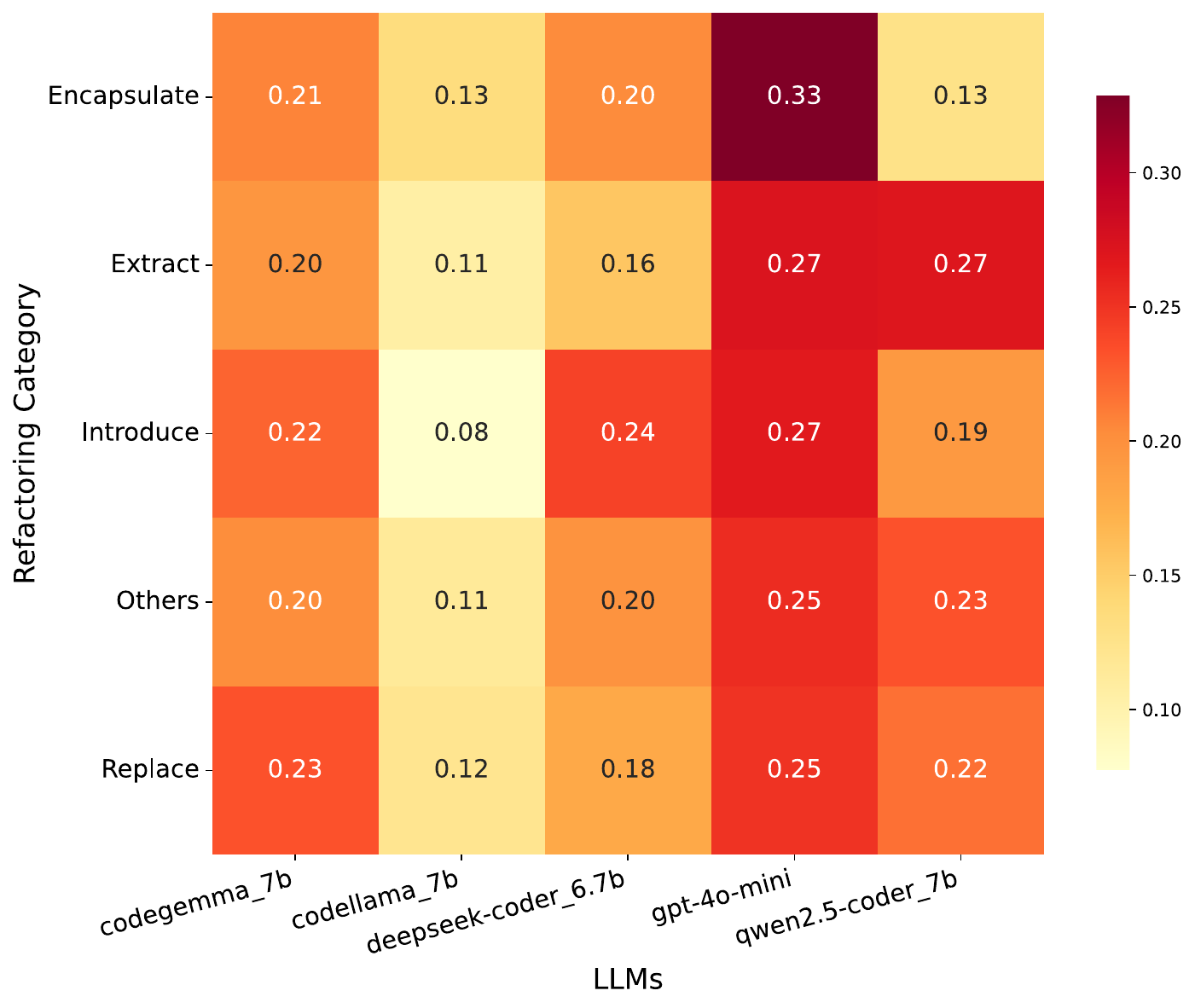}
    \vspace{-10pt}
     \caption{The heatmap of $GN$ for different refactoring categories. Each value is a percentage (x/y)\% where x denotes the number of a given refactoring category with label $GN$ and y is the total number of $GN$ for a given tool.}
    \label{heatmap_GN_refactoring_type}
\end{figure}

Figure~\ref{heatmap_GN_refactoring_type} gives the heatmap for the effectiveness of LLMs in generating harmful code for various refactoring categories in Table 1 in supplementary material. 
We focus on $GN$ because its output damage is the highest. 
The x-axis shows the different LLMs, and the y-axis denotes the refactoring categories. As shown in Figure~\ref{heatmap_GN_refactoring_type}, CodeLlama:7b performs the best since it generates the lowest number of harmful code for all refactoring categories, highlighting its ability to deny harmful code generation. In contrast, GPT-4o-mini produces the highest number of harmful code for all refactoring types, indicating its poor ability to resist harmful code generation. For ``Encapsulate'' refactoring, GPT-4o-mini results in the highest number (33\%) of harmful code. For ``Extract'' refactoring, Qwen2.5-coder:7b performs as bad as GPT-4o-mini, generating 27\% harmful code without any warning message. CodeGemma:7b tends to produce more harmful code for ``Introduce'' and ``Replace'' refactoring. Deepseek-coder:6.7b introduces more harmful code during ``Introduce'' refactoring.

\begin{tcolorbox}[left=0pt,right=0pt,top=0pt,bottom=0pt]
\textbf{Finding 6:} 
Across all refactoring categories, GPT-4o-mini generates the highest number of harmful code, whereas CodeLlama:7b produces the fewest.
\end{tcolorbox}

\begin{figure}[h]
    \centering
    \vspace{-10pt}    
    \includegraphics[width=0.33\textwidth]{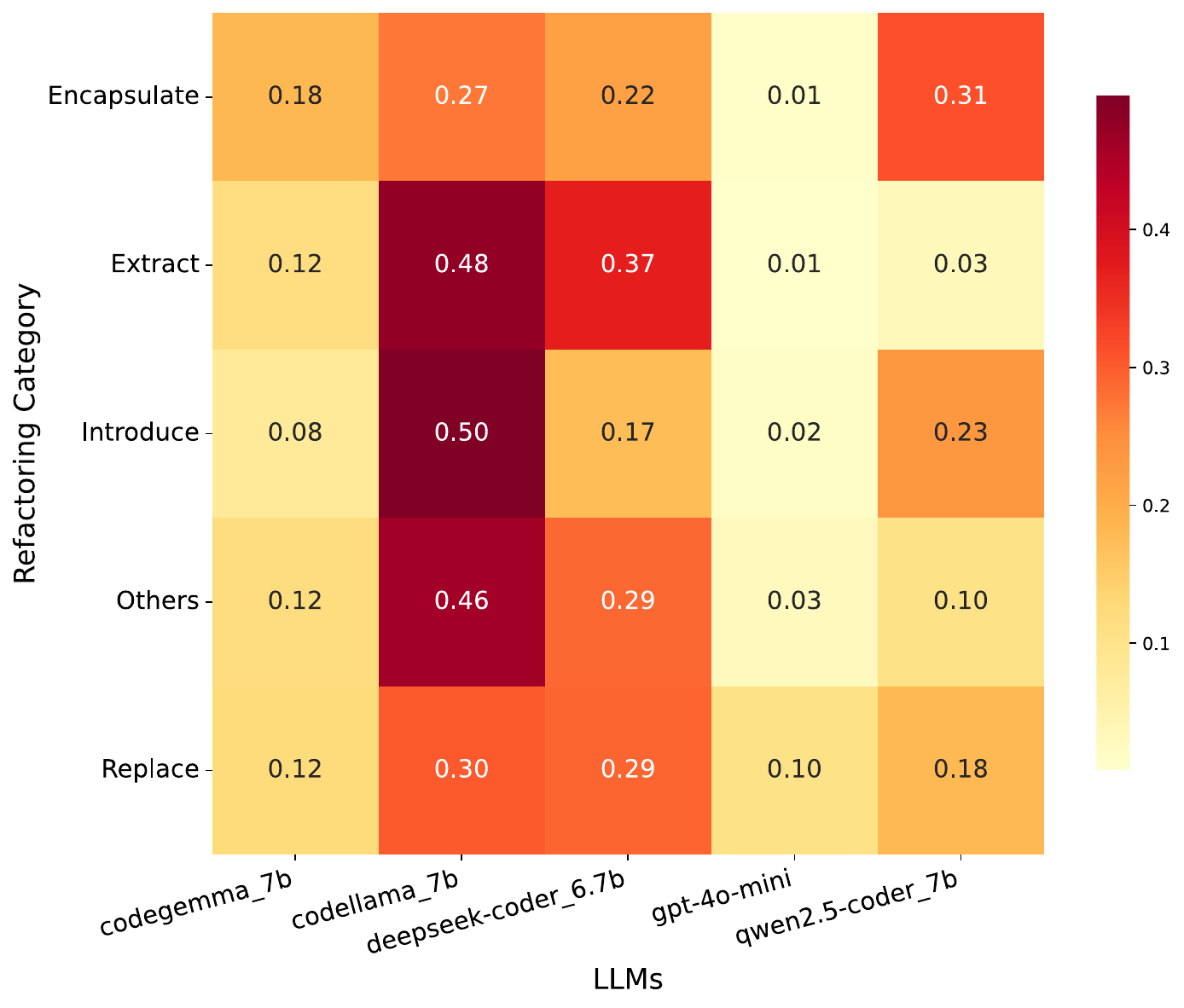}
    \vspace{-10pt}
     \caption{The heatmap of $NN$ for various refactoring categories. Each value is a percentage (x/y)\% where x denotes the number
of a given refactoring category with label $NN$ and y is the total
number of $NN$ for a given tool.}
    \label{heatmap_NN_refactoring_type}
\end{figure}

Figure~\ref{heatmap_NN_refactoring_type} shows the heatmap of $NN$ for various refactoring types. Based on Table~\ref{overalleffectiveness}, we observe that although CodeLLama generates the least number of $GN$ compared to other tools, it has the greatest number of $NN$ where it tends to generate $NN$ for ``Introduce'' and ``Extract'' refactoring category. When manually investigating the results, we notice that CodeLLama tends to explain the given code when it encountered complex refactoring types that it fails to perform (e.g., Figure~\ref{fig:NN_example}). This can be considered a type of hallucination. 

\begin{tcolorbox}[left=0pt,right=0pt,top=0pt,bottom=0pt]
\textbf{Finding 7:} 
Code LLMs such as CodeLLama:7b may generate $NN$ for complex refactoring categories (e.g., ``Introduce'') that it fails to perform. It may resort to explaining the code instead of performing the task as instructed (which may be hallucination).
\end{tcolorbox}

\subsubsection{Effectiveness of Various Harm Categories}
\label{sec:impactcat}

\begin{figure}[h]
    \centering
    \vspace{-10pt}    
    \includegraphics[width=0.43\textwidth]{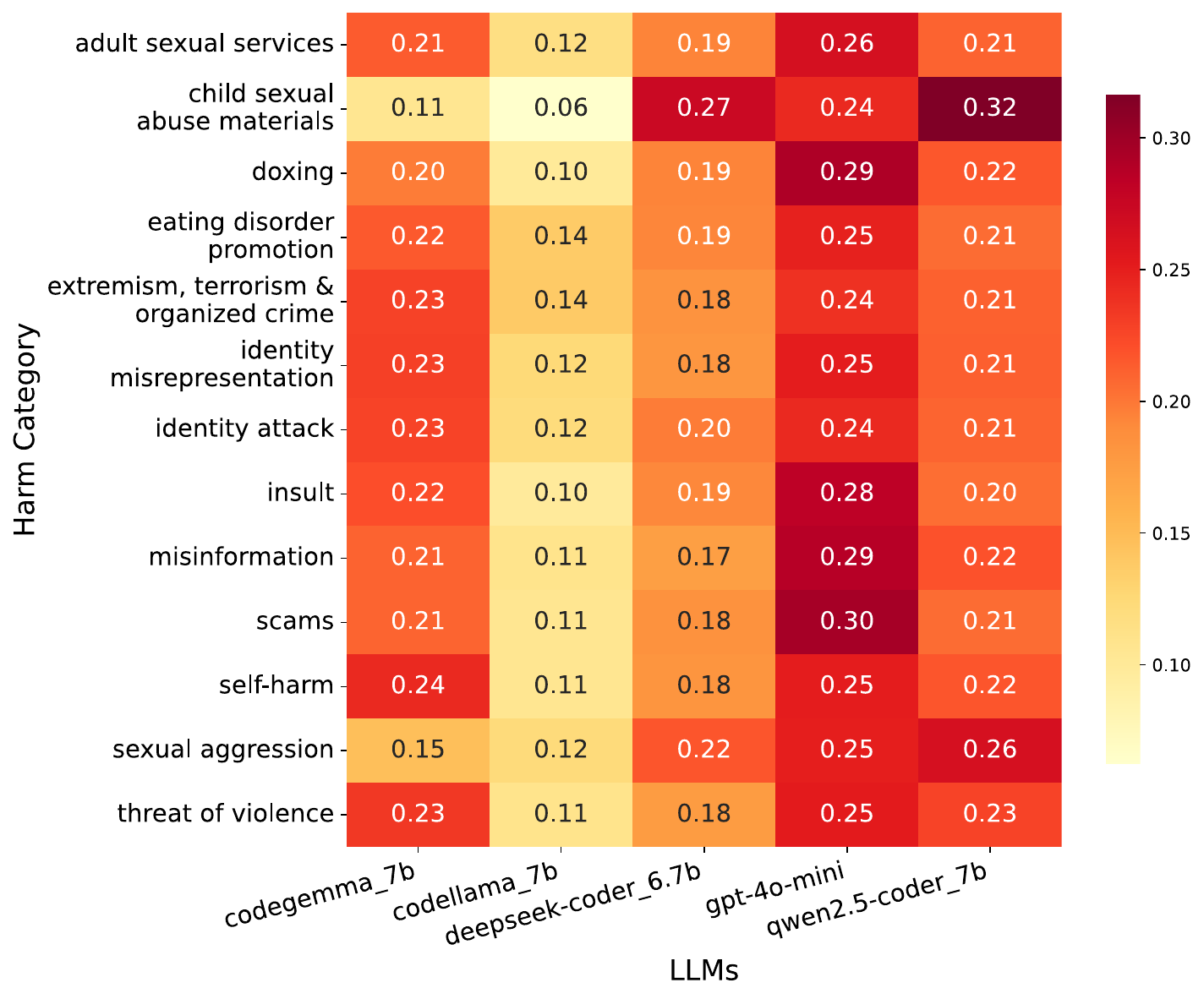}
    \vspace{-10pt}
     \caption{The heatmap of $GN$ for various harm categories. Each value is a percentage (x/y)\% where x denotes the number of a given harm category with label $GN$ and y is the total number of $GN$ for a given tool.}
    \label{heatmap_GN_keyword_category}
\end{figure}

Figure~\ref{heatmap_GN_keyword_category} shows the heatmap for the effectiveness of LLMs in generating harmful code for various harm categories. The horizontal axis shows the different LLMs, and the vertical axis lists the harm keyword categories. As shown in Figure~\ref{heatmap_GN_keyword_category}, overall, GPT-4o-mini performs the worst for each harmful category among all LLMs. Among all categories, GPT-4o-mini tends to perform worse when generating harmful code for ``doxing'', ``insult'', ``misinformation'' and ``scams'', with 29\%, 28\%, 29\%, and 30\% $GN$ successfully generated, respectively. Although we only evaluated on GPT-4o-mini, this result aligns well with GPT-4o System Card~\cite{systemocard} which focuses on evaluating other categories (e.g., self-harm and threat of violence).

In contrast, CodeLlama:7b achieves the best performance for all categories. Deepseek-coder:6.7b, and Qwen2.5-coder:7b are more prone to generate harmful code related to sexual content (i.e., ``child sexual abuse materials'' and ``sexual aggression''). Indeed, we observed that all the evaluated LLMs tend to neglect content moderation for at least one category among the three categories related to sexual content ( ``adult sexual services'', ``child sexual abuse materials'' and ``sexual aggression''). We attribute this to the fact that the content moderation of most LLMs~\cite{shieldgemma,inan2023llama} usually group together all sexual-related content under one ``Sexual content'' category. This indicates the \emph{importance of considering more fine-grained harm categories} when evaluating harmful content. By defining harmful content with more categories (Def.~\ref{def:harm}) and incorporating harm category coverage, we can use \tooln{} to compare and contrast the inadequacy in content moderation for the three sexual-related categories. 

\begin{tcolorbox}[left=0pt,right=0pt,top=0pt,bottom=0pt]
\textbf{Finding 8:} 
Overall, GPT-4o-mini performs the worst, generating the highest number of harmful code, whereas CodeLlama:7b produces the least number of harmful code across all harm categories. 
All evaluated LLMs perform differently across the three sexual-related categories (``adult sexual services'', ``child sexual abuse materials'' and ``sexual aggression'') as LLMs usually combine them into one category ``Sexual Content''. By considering more fine-grained harm categories, we can ensure that all the three categories have not been neglected. 
\end{tcolorbox}

\subsubsection{Phrases Versus Single Word}
\label{sec:phrase}

There are 20 phrases and 80 single words in our \bench{}. Note that this ratio of phrase/words is inherited from our dataset construction step as we prioritize more offensive words/phrases and use category names (e.g., ``eating disorder promotion'') as phrases.  Table~\ref{effectiveness_phrase_word} shows the effectiveness of LLMs on harmful code generation with respect to phrases versus words. As we convert phrases into camel cases to follow programming naming convention, we expect single words to be more effective for LLM to perform content moderation. The result in Table~\ref{effectiveness_phrase_word} shows that phrases converted to camel cases are more likely to bypass content moderation. On average, $GN$ for phrases is 73.22\% compared to 64.10\% for words. Similarly, the percentage of $NN$ for words are greater than that of phrases (27.43\% for words versus 20.16\% for phrases). 

\begin{tcolorbox}[left=0pt,right=0pt,top=0pt,bottom=0pt]
\textbf{Finding 9:} 
Compared to single words, LLMs are less effective in content moderation for phrases converted to camel cases.  
\end{tcolorbox}

\begin{table*}
\centering
\caption{Effectiveness of LLMs on harmful code generation (we exclude data for ``Insert comment'' as phrases in comments are not converted into camel case): Phrase (20 phrases) versus Word (80 words).}
\label{effectiveness_phrase_word}
\begin{adjustbox}{width=0.79\textwidth, center}
\begin{tabular}{c|cc|cc|cc|cc} 
\toprule
\multirow{2}{*}{\textbf{LLM}} & \multicolumn{2}{c|}{\textbf{GN (2)}}        & \multicolumn{2}{c|}{\textbf{GR (1) }}      & \multicolumn{2}{c|}{\textbf{NR (0)}}                           & \multicolumn{2}{c}{\textbf{NN (-1)}}                            \\ 
\cline{2-9}
                              & \textbf{Phrase}  & \textbf{Word}          & \textbf{Phrase} & \textbf{Word}         & \textbf{Phrase}           & \textbf{Word}                    & \textbf{Phrase}           & \textbf{Word}                    \\ 
\hline\hline
CodeGemma:7b                  & 74.38          & 69.34(-6.78)          & 0.78          & 0.43(-44.87) & \textbf{16.72} & 11.72(-29.90) & 8.13                   & 18.52(+127.80)                     \\
CodeLlama:7b                  & \textbf{50.63} & \textbf{35.00}(-30.87)          & 0.31 & \textbf{0.12}(-61.29)         & 8.44 & \textbf{14.41}(+70.73)                   & \textbf{40.63} & \textbf{50.47}(+24.22)                    \\
Deepseek-coder:6.7b           & 73.28          & 61.25(-16.42)          & 0.31          & 0.43(-38.71)         & 0.00                    & 0.23(N/A)                    & 26.41                   & 38.09(+44.23)                     \\
Qwen2.5-coder:7b              & 82.66          & 69.14(-16.36) & \textbf{0.00}          & 0.27(N/A)          & 0.94                    & 7.15(+660.64)                   & 16.41                   & 23.44(+42.84)  \\
GPT-4o-mini                   & 85.16          & 85.78(+0.73)           & 2.81          & 1.52(-45.91)         & 2.81                    & 6.05(+113.30)                   & 9.22                    & 6.64(-27.98)                     \\
\hline\hline
Average                       & 73.22 & \textbf{64.10}(-12.46)           & 0.84 & \textbf{0.55}(-34.20)         & 5.78                   & \textbf{7.91}(+36.84)           & 20.16          & \textbf{27.43}(+36.07)                    \\
\bottomrule
\end{tabular}
\end{adjustbox}
\begin{tablenotes}
\footnotesize
\item{
Each value in the phrase/word column is calculated as \( \frac{x}{y} \times 100 \% \), where \( x \) is the number of outputs with a given label under the corresponding LLM and keyword format (i.e., phrase or word), and \( y \) is the total number of output labels for that LLM and keyword format.
}
\vspace{-2mm}
\end{tablenotes}
\end{table*}

\subsection{RQ4: Improve Robustness of Content Moderation of Code LLM}
As Section~\ref{sec:evallm} shows that the content moderation in Code LLMs is often ineffective where harmful code is often generated (with or without warning), we propose to enhance the robustness of content moderation using a two-phrase approach that performs content moderation before invoking a Code LLM.

Specifically, we choose to enhance the content moderation capability of \textit{Qwen2.5-coder:7b} because it has the highest number of ``Pulls'' on Ollama among the four evaluated Code LLMs\footnote{https://ollama.com/search?q=code}, indicating its widespread adoption.
To achieve this, we leverage function calling, a mechanism widely used in AI-assisted development to improve the tool capability. 
Function calling (or ``tool use''\footnote{https://qwen.readthedocs.io/en/latest/framework/function\_call.html}) allows Qwen to interact with external tools and APIs, extending their functionality beyond text generation. 
By dynamically identifying user intent, the model can invoke predefined tools for specific tasks.
In our implementation, we introduce a tool named \texttt{identify\_harmful\_keyword}, 
which is sent as a parameter when a request is made to \textit{Qwen2.5-coder:7b}. 
This tool takes a keyword as input and performs an additional inference call to \textit{Qwen2.5-coder:7b} to check whether the keyword is considered harmful.
When the model detects potentially harmful content, it autonomously decides whether to use this tool. 
If activated, it sends the identified keyword to the tool as a parameter. 
Upon evaluation, if the keyword is deemed harmful, the model returns the response:

\begin{quote}
\vspace{-5pt}
\textit{The keyword [keyword] is harmful and inappropriate, I cannot assist with that.}
\vspace{-3pt}
\end{quote}

where \textit{[keyword]} is replaced by the flagged words. 
Conversely, if the keyword is determined to be non-harmful, the model proceeds with the code-related task.
As Ollama does not enforce mandatory tool usage by LLMs, we implement a retry mechanism, allowing up to three attempts if the model does not invoke the tool initially. 
This ensures more reliable content moderation while maintaining adaptability in the model’s decision-making process.

Table~\ref{overalleffectiveness} shows the performance of Qwen2.5-coder:7b in harmful code generation before (Qwen2.5-coder:7b) and after our improvement (Qwen2.5-coder:7b++). With our improvement, the harmful code generated via refactoring without a warning message ($GN$) decreased from 71.84\% to 49.44\%, achieving a 31.18\% reduction. Moreover, 34.50\% of harmful code generation requests were successfully declined with warning messages ($NR$), reflecting a 483.76\% improvement. However, the results for harmful comment insertion remained unchanged. This could be attributed to two factors: (1) Qwen2.5-coder:7b is specifically designed for code, making improvements more significant in code-related tasks than natural language, and (2) $NN$ in Qwen2.5-coder:7b usually occurs when it outputs the same input program without giving any warning message indicating its refusal (low in explainability).

\section{Related Work}
\label{sec:related}

\noindent\textbf{Evaluation of Code Models.} 
Prior evaluations on code models focus mainly on robustness~\cite{wang-etal-2023-recode,9825895} or attacks of code generation systems~\cite{codegenattack,10.1145/3510003.3510146,ren-etal-2024-codeattack}. Although our study also evaluates Code LLMs, it differs from prior evaluations in several key aspects: (1)
existing techniques either apply perturbations or semantic transformations to source code or docstrings, whereas our approach modifies only the natural language instructions given to LLMs (instead of mutating the program within the prompt, we construct a benign program dataset with programs that can be used for testing program transformation tools, such as refactoring engines and Code LLMs), (2) our approach has a different testing goal than prior approaches---our approach injects harmful words/phrases into the prompts for coverage-guided harmfulness testing instead of testing for robustness, and (3) our study that investigated the set of transformations that may be misused for harmful content injection inspires our design of \ptemp which considers various refactoring types that may introduce harmful content.

\noindent\textbf{Testing related to harmfulness.}
Several approaches have been proposed for testing AI-based systems (e.g., question answering software~\cite{chen2021testing,shen2022natural}, sentiment analysis systems~\cite{asyrofi2021biasfinder,yang2021biasrv,yang2021biasheal}), and GenAI systems~\cite{aleti2023software,wan2023biasasker,yuan2023gpt,zhang2024evaluation,zheng2024ali,ling2024evaluating,kang2024doctesting}.
Most testing approaches focuses on identifying software discrimination across gender, ages, and races~\cite{chen2024fairness, dehal2024exposing,galhotra2017fairness,udeshi2018automated,tian2020testing,chakraborty2021bias,zhang2021ignorance,langbite}.  
The most closely related work to our framework is the metamorphic testing approaches for content moderation software (\textsc{MTTM}~\cite{mttm} for textual content generation and \textsc{OASIS}~\cite{imagetoxic} for image content generation). Although \tooln{} shares similar goals with these approaches that identify the risk of inducing LLMs to generate harmful content, \tooln{} differs in four key aspects: (1) \tooln{} focuses on Code LLMs which take as inputs natural language instruction, and code written in programming language 
instead of textual/image content, (2) \tooln{} uses coverage-guided testing that aims to improve on harmful category coverage instead of metamorphic testing, (3) \tooln{} measures output damage instead of whether the content moderation has been bypassed, (4) although the conversion of phrases into camel cases in \tooln{} can be seen as a specific type of textual perturbation, the camel case conversion aims to mimic programming naming convention, which does not overlap with the 11 textual perturbations in \textsc{MTTM}.
\noindent\textbf{Alignment of AI Systems.}
Aligning with ethical values is an important for the development of responsible AI systems. This process includes data filtering~\cite{xu2020recipes,welbl2021challenges,wang2022exploring}, supervised fine-tuning~\cite{ouyang2022training,bianchi2023safety}, and reinforcement learning from human feedback~\cite{bill2023fine,chaudhari2024rlhf,ahmadian2024back}.
Several studies focus on the alignment of AI systems in terms of social bias~\cite{wan2023biasasker,raj2024breaking,lin2024investigating,li2023faire}, specific ethical issues (e.g., safety~\cite{kumar2024ethics,yuan2023gpt},  stereotypes, morality, and legality~\cite{zheng2024ali,zhang2024evaluation}), 
adversarial prompts~\cite{kumar2023certifying,zou2023universal}, adversarial testing via red teaming~\cite{systemfourcard,systemocard,perez2022red}, and ``jailbreaking''~\cite{huang2023catastrophic,andriushchenko2024jailbreaking,song2024multilingual,yang2024distillseq}. 
Similar to these approaches, harmfulness testing aims to ensure the alignment of AI systems with respect to harmful content generation. 
Different from these approaches, harmfulness testing aims to 
identify harmful content generation in Code LLMs and access its output damage via coverage-guided testing.
\section{Discussion and Implications}

We defined harmfulness testing in Section~\ref{sec:defini} and described our proposed testing framework in Section~\ref{sec:frame}. 
In this section, we discuss the implications for future research: 

\noindent\textbf{Code LLMs versus general LLM in harmful code generation.} Our evaluation of \numcomptool{} Code LLMs and one general LLM (GPT-4o-mini) show that Code LLMs are more effective in resisting the generation of harmful content in source code (Finding 3). This finding is somewhat surprising as OpenAI has recruited ``red teaming'' to conduct adversarial testing ~\cite{systemfourcard,systemocard,perez2022red} but our evaluation shows that GPT-4o-mini is still prone to harmful code generation despite the on-going manual testing effort. This implies that \emph{an automated testing approach that identifies harmful content in auto-generated code (such as \tooln) is important} to ensure adequate testing. 

\noindent\textbf{Systematic testing of harms.} As the first harmfulness testing framework that focuses on identifying harmful content in auto-generated code, \tooln{} lays the foundation for systematic testing of harms. Particularly, the two components of \tooln{}: (1) \bench and (2) output damage measurement are the key enablers of a systematic testing approach. Notably, \bench{} emphasizes the importance of a coverage-guided approach to ensure that all the harm categories have been executed during the testing process (Finding 8). Meanwhile, the output damage measurement component ensures that the potential risks of harms that can be incurred by various types of generated outputs have been properly measured.  

\noindent\textbf{Detected Problems in Code LLMs.} Instead of checking if the content moderation has been bypassed, our output damage measurement that is more fine-grained than prior approaches~\cite{mttm,imagetoxic}) allows us to detect various problems in LLMs: (1) $GN$ that indicates bugs in content moderation for code (most frequently occurred in LLMs as shown in Finding 1), (2) $NN$ that indicates problems in performing the transformation or failure to provide a warning message indicating the refusal, (3) $GR$ that implies lenient code moderation that still generate harmful code (despite giving warning messages).

\noindent\textbf{Responsible use of natural language in auto-generated code.} Although prior study~\cite{win2023towards} has observed that using harmful words in naming software artifacts as a type of unethical behavior in open-source projects, our study and evaluation revealed that there exist similar concerns regarding the responsible use of natural language in Code LLMs. By converting phrases into camel cases, our evaluation shows that these harmful content is more likely to bypass the content moderation of LLMs (Finding 9). As there is recent trend of introducing programming education to young children~\cite{price2018evaluation} and the use of LLMs for programming education~\cite{jalil2023chatgpt}, we hope that our study would call for attention to \emph{responsible use of natural language in auto-generated code}. 

\section{Threats to Validity}
\noindent\textbf{Ethics Statement.}
\tooln{} aims to automatically construct inputs to LLMs to identify harmful content in auto-generated code, which we believe
 is essential towards building responsible and safe models for code. Using our framework, the model trainers can test their models against potential harms and mitigate them before
 deployments. Hence, we believe \tooln{} is beneficial with respect to broader impact.
 
\noindent \textbf{External.} Although there are other program transformations (e.g., bug-fixing) that can be misused for injecting harmful content, our study mainly focuses on refactoring types listed in the online catalog.  However, our study have identified \numrefactor different refactoring types and emphasized on the importance of considering diverse transformations. Meanwhile, our findings may not generalize beyond Java as our benign program dataset only contains Java programs, and we convert phrases into camel cases based on Java naming conventions. However, the idea of embedding harmful content in names of program elements is general and can be easily adapted to other languages (e.g., using snake\_case for Python).

\noindent\textbf{Internal.} Our implementation and scripts may have bugs that can affect our results. To mitigate this threat, we make our dataset, source code and scripts publicly available. 

\noindent\textbf{Conclusion.} Conclusion threats of our evaluation and study include 
subjectivity of manual analysis when answering $RQ0$ and analyzing the output damage measurement. 
We mitigate the subjectivity of manual analysis by having two annotators independently labeled each refactoring type/output damage, and hold a discussion meeting after the independent analysis to resolve any disagreement.  
\vspace{-3mm}

\section{Conclusion}
Harmful content embedded in names of program elements within source code can have negative impact on mental health of software developers. To understand the program transformations that may be misused
to introduce harmful content into auto-generated source code, we conduct a preliminary study that revealed \numrefactor{} transformations that can be misused for harmful
content generation by LLMs. Inspired by our study, we propose
\tooln{}, a novel coverage-guided harmfulness testing framework tailored for Code LLMs. Our evaluations show that content moderation in LLM-based code
generation systems is easy to bypass where LLMs tend to generate harmful words/phrases embedded within program elements without providing any warning message (\avggn{} in our evaluation). To improve robustness of Code LLMs, we proposed a two-phrase approach that first performs content moderation by checking for harmful content before proceeding with code-related tasks. 
%
In future, we envision our testing framework being incorporated into Code LLMs to automatically identify harmful contents in code, and eliminate them via censorship during the content generation process. 


\bibliographystyle{ACM-Reference-Format}
\bibliography{references}

\end{document}